\documentclass{PoS}

\usepackage[version=4]{mhchem}
\usepackage[table]{xcolor}
\usepackage{multirow}
\usepackage{physics}
\usepackage{xspace}

\def \molgyr {molecules~g$^{-1}$~yr$^{-1}$\xspace}
\newcommand{\model}[1]{\mathcal{M}_{#1}}

\title{Obtaining a History of the Flux of Cosmic Rays using In
Situ Cosmogenic \ce{^14C} Trapped in Polar Ice}

\ShortTitle{Cosmic Ray Flux and Carbon-14}

\author{\speaker{Segev BenZvi}$^{1}$, Vasilii V. Petrenko$^{2}$, Benjamin Hmiel$^{2}$, Michael Dyonisius$^{2}$, Andrew M. Smith$^{3}$, Bin Yang$^{3}$, and Quan Hua$^{3}$ \\
$^{1}$Department of Physics and Astronomy, University of Rochester, Rochester, NY, USA\\
$^{2}$Department of Earth and Environmental Sciences, University of Rochester, Rochester, NY, USA\\
$^{3}$Australian Nuclear Science and Technology Organisation (ANSTO), Locked Bag 2001, Kirrawee DC, NSW 2232, Australia\\
E-mail: \email{sbenzvi@ur.rochester.edu}
}

\abstract{
Carbon-14 (\ce{^14C}) is produced in the atmosphere when neutrons from cosmic-ray air showers are captured by \ce{^14N} nuclei. Atmospheric \ce{^14C} becomes trapped in air bubbles in polar ice as compacted snow (firn) transforms into ice. \ce{^14C} is also produced {\sl in situ} in ice grains by penetrating cosmic-ray neutrons and muons. Recent ice core measurements indicate that in the \ce{^14CO} phase, the \ce{^14C} is dominated by the {\sl in situ} cosmogenic component at most ice coring sites. Thus, it should be possible to use ice-bound \ce{^14CO} to reconstruct the historical flux of cosmic rays at Earth, without the transport and deposition uncertainties associated with \ce{^10Be} or the carbon cycle uncertainties affecting atmospheric \ce{^14CO2}. The measurements will be sensitive to the cosmic-ray flux above the energy range most affected by solar modulation. We present estimates of the expected sensitivity of \ce{^14CO} in ice cores to the historical flux of Galactic cosmic rays, based on recent studies of \ce{^14CO} in polar ice.
}

\FullConference{36th International Cosmic Ray Conference -ICRC2019-\\
		July 24th - August 1st, 2019\\
		Madison, WI, U.S.A.}

\begin{document}

\section{Introduction and Motivation}\label{sec:intro}

Cosmogenic radionuclides are a long-established tool for determining the ages of biological samples \cite{Libby:1946}, for the study of long-term changes in Earth's climate \cite{Miyake:2012}, for monitoring the history of solar irradiance \cite{Usoskin:2013, Miyake:2017, Wu:2018}, to trace the historical flux of Galactic cosmic rays \cite{Wieler:2011}, and to look for evidence of supernovae in the vicinity of the solar system \cite{Ellis:1995qb, Fields:1998hd, Brakenridge:2011, Wallner:2016}. The isotopes most famously used for dating, \ce{^10Be} and \ce{^14C} in tree rings and ice cores, are produced by the interaction of secondary particles from low-energy Galactic cosmic rays with oxygen and nitrogen in Earth's atmosphere. ``Low-energy'' in this context refers to primary cosmic rays between 10 and 100 GeV per nucleon, above the geomagnetic cutoff at most latitudes but below the energies where solar modulation ceases to have a major impact on the cosmic-ray flux at Earth. Thus, changes in the accumulation rate of these isotopes are useful for studying the past behavior of the heliosphere. 

There is evidence in records of \ce{^10Be} and \ce{^14C} to indicate significant excursions from constant production rates in the atmosphere. These excursions can be slowly varying over thousands of years, or impulsive changes lasting from several years to several decades. It is typical to assume that time-varying production rates are due to geomagnetic effects and the effects of the heliosphere on the Galactic cosmic rays \cite{Usoskin:2013, Wu:2018, Knudsen:2009}, though it has been suggested that gamma rays from nearby supernovae could increase the atmospheric production of \ce{^10Be}, \ce{^14C}, and \ce{^36Cl} on time scales of decades \cite{Brakenridge:2019ram}.

Studies of the heliosphere using atmospheric radionuclides rest on the assumption that the flux of Galactic cosmic rays is constant over the time scales of interest. At some level this must not be correct --- the flux of cosmic rays is affected by the acceleration of charged particles in supernovae and supernova remnants, the passage of the solar system through local interstellar bubbles, the long-term motion of the solar system through the spiral arms of the Milky Way, and very long-term variations in the Galactic star formation rate \cite{Scherer:2006, Frisch:2010ud, Frisch:2018ult}. Meteorites can be used to study the variability of the cosmic ray flux over timescales of order several years to $10^6$ years, since they accumulate cosmogenic radionuclides with a wide range of lifetimes.
Estimates of radionuclide production in meteorites over these different time intervals suggest the Galactic cosmic-ray flux has been stable during the past $\sim10^6$~yr, though the estimates are affected by considerable systematic uncertainties due to solar modulation, meteoroid orbits, and the shielding effects of the meteoroid surfaces. Current estimates of a constant historical cosmic-ray flux could be uncertain at the level of at least 30\% \cite{Wieler:2011}.

In this work, we discuss a new use of \ce{^14C} in polar glacial ice to test the variability in the flux of high-energy cosmic rays on time scales $\sim10^4$~yr. \ce{^14C} in polar ice is produced {\sl in situ} by muons originating from cosmic rays above 100 GeV, in principle recording variations in Galactic cosmic ray flux only, without significant influences from geomagnetic or heliospheric field variations. In Section~\ref{sec:c14}, we discuss the production and accumulation of \ce{^14C} in deep ice and its potential use as a tracer of the historical cosmic-ray flux. In Section~\ref{sec:sensitivity} we estimate the sensitivity of \ce{^14C} profiles in ice cores to several different variable flux scenarios. Section~\ref{sec:conclusion} contains a brief discussion and conclusion.

\section{Carbon-14 Accumulation and Production in Deep Ice}\label{sec:c14}

\begin{figure}[ht]
  \centering
  \includegraphics[width=\textwidth]{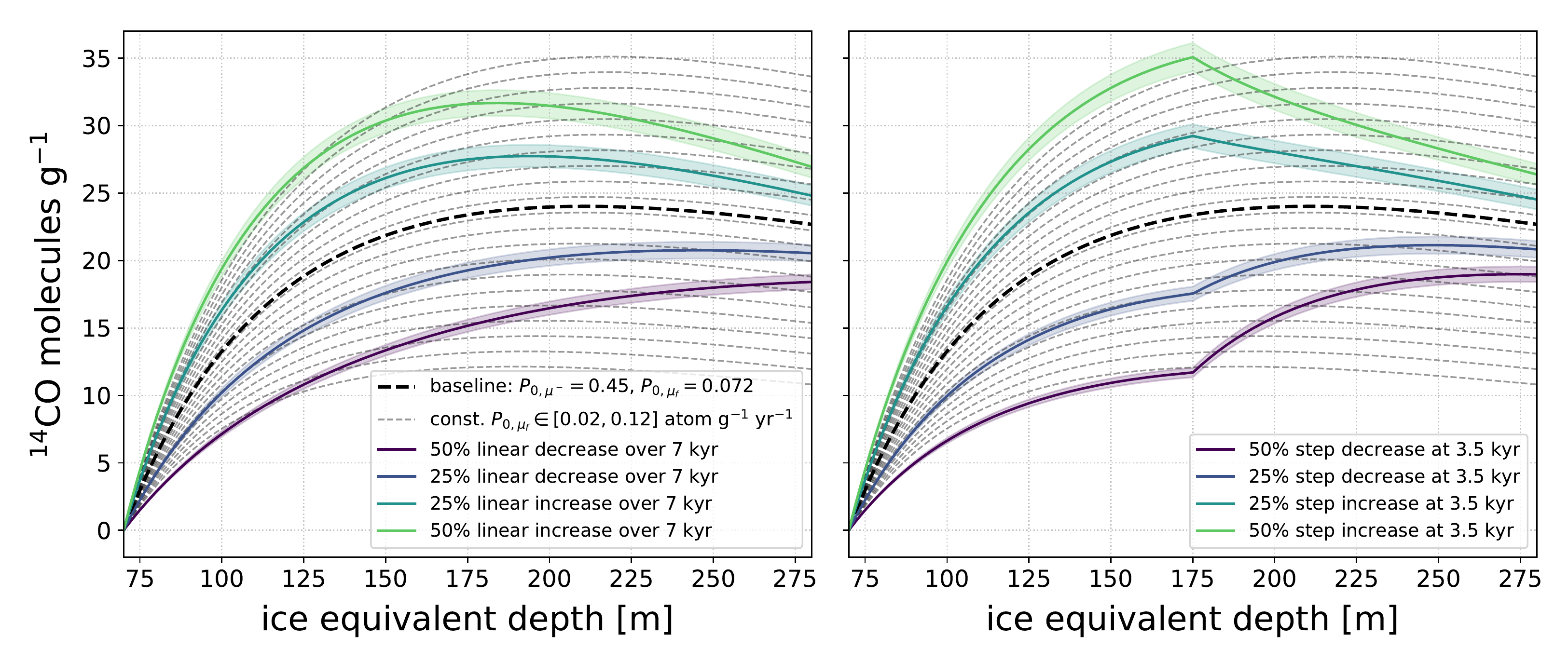}
  \caption{Simulated \ce{^14CO} profiles in Dome C ice based on surface production rates measured at Taylor Glacier (thick dashed curve labeled ``baseline''). The thin dashed lines correspond to simulations in which the production rate due to fast muons, $P_{0,\mu_f}$, is varied between $0.02$ and $0.12$~molecules~g$^{-1}$~yr$^{-1}$ (a much larger range than needed to explain data from Taylor Glacier). For comparison, the solid lines show calculations in which the production rates due to slow and fast muons are allowed to vary in time, with linear changes shown at left and step-like changes shown at right. The bands indicate conservative $1\sigma$ measurement uncertainties, assumed to be 3\%. Trapped atmospheric \ce{^14CO} is expected to be $\sim1.5$ molecules~g$^{-1}$ and is ignored.}
  \label{fig:co14_profile_models}
\end{figure}

Carbon-14 in polar glacial ice comes from two sources: trapping of air, which contains \ce{^14C} in its methane, carbon monoxide, and carbon dioxide phases; and {\sl in situ} cosmogenic production. Atmospheric \ce{^14C} is created when thermal neutrons ($E_n\lesssim1$~eV) from cosmic ray air showers are captured by \ce{^14N}. This \ce{^14C} is quickly oxidized to \ce{^14CO}, then more slowly to \ce{^14CO_2}, after which it can enter the biospheric carbon cycle, also resulting in some \ce{^14CH_4} emission to the atmosphere.
{\sl In situ} production of \ce{^14C}, on the other hand, occurs via spallation of \ce{^16O} by fast neutrons ($E_n\sim1$~MeV) in the first few meters of the firn layer; capture of slow $\mu^-$ particles by \ce{^16O}, most important in the top $\sim40$~m of firn; and interactions with ``fast'' muons ($E_\mu\gtrsim10$~GeV) which can penetrate to greater depths \cite{Petrenko:2016}.

Preliminary measurements of the firn open porosity, firn matrix, and ice cores to 130~m depth at Greenland Summit \cite{Hmiel:2019} are showing that most of the {\sl in situ} cosmogenic \ce{^14CO} is lost from the firn grains immediately after production, with further loss via a slower leakage process. Most of the leaked \ce{^14CO} then escapes to the atmosphere. Substantial accumulation of {\sl in situ} \ce{^14CO} only begins below the lock-in depth, where the first impermeable ice layers form and the firn porosity is effectively sealed off from the atmosphere. The accumulated {\sl in situ} \ce{^14CO} in deeper ice thus originates almost entirely from production by deep-penetrating muons that originate from high-energy primary cosmic rays ($>100$~GeV per nucleon).

Further preliminary measurements of \ce{^14CO} in ablating ice at Taylor Glacier, Antarctica, provide constraints on \ce{^14CO} {\sl in situ} production rates by slow and fast muons \cite{Dyonisius:2019}. Adapting a model of \ce{^10Be} and \ce{^26Al} production in rock by muons from Balco {\sl et al.} \cite{Balco:2008}, the surface production rates of $\ce{^14C}$ due to slow and fast muons interacting in ice are estimated to be $P_{0,\mu^-}=0.46\pm0.03$~\molgyr and $P_{0,\mu_f}=0.071\pm0.020$~\molgyr, respectively. Combining the information on \ce{^14CO} retention in the firn from Greenland Summit with \ce{^14CO} production rates from Taylor Glacier, we can estimate the expected {\sl in situ} \ce{^14CO} concentration profiles in deep ice at other locations. At most ice coring sites, the {\sl in situ} cosmogenic \ce{^14CO} component is predicted to be much larger than $\sim1.5$ \ce{^14CO} molecules g$^{-1}$ due to trapped air. Thus, measurements of \ce{^14CO} from deep ice cores can in principle be used to examine the variability in the flux of high-energy cosmic rays.

An excellent site for such a study would be Dome C in Antarctica. Dome C has a very low and stable accumulation rate of 3~cm ice equivalent yr$^{-1}$, maximizing exposure to cosmic rays and accumulation of {\sl in situ} \ce{^14CO}, and has good infrastructure for drilling of ice cores. Logistically uncomplicated dry-drilled 300~m deep ice cores from Dome C can capture the cosmic-ray history over a period of roughly 7 kyr, covering most of the Holocene epoch.

Simulated \ce{^14CO} profiles at Dome C are shown in Fig.~\ref{fig:co14_profile_models}. To produce the plots, we used muon fluxes following Balco {\sl et al.} \cite{Balco:2008} and started with constant muon \ce{^14CO} production rates $P_{0,\mu^-}$ and $P_{0,\mu_f}$ as determined for Taylor Glacier. We then varied these (temporally constant) production rates within uncertainties, and also examined several scenarios for temporal variations in production rates. We note that at Dome C, we expect substantial accumulation of {\sl in situ} cosmogenic \ce{^14CO} to only begin at the lock-in depth in deep firn ($\sim70$~m ice equivalent depth). While the rate of \ce{^14CO} production by muons declines with depth, it is significant for the entire depth range shown in Fig.~\ref{fig:co14_profile_models}. This effectively causes the ice core \ce{^14CO} record of past production rates to be highly smoothed in time. In the following section, we estimate the sensitivity of the depth profile of \ce{^14CO} to temporal changes in the production rates, which should be directly related to changes in the flux of cosmic rays above 100~GeV. 

\section{Sensitivity to Changes in the Production Rate of Carbon-14}\label{sec:sensitivity}

We wish to compute the sensitivity of measurements of locked-in \ce{^14CO} at Dome C to variations in the flux of high-energy cosmic rays. From the computed profiles shown in Fig.~\ref{fig:co14_profile_models} it is clear that large deviations from the ``baseline'' profile of \ce{^14CO} can be produced by time-varying fluxes or by constant fluxes with very large or small $P_{0,\mu_f}$. Thus, the analysis must be sensitive to the shape of the \ce{^14CO} profile and not its normalization.

To discriminate time-varying models from constant-flux models, we use the Bayes Factor
\begin{align}\label{eq:bf}
  B_{01} &= \frac{\Pr(\model{0}|\ce{^14CO})}{\Pr(\model{1}|\ce{^14CO})}
          = \frac{\Pr(\ce{^14CO}|\model{0})}{\Pr(\ce{^14CO}|\model{1})} \cdot
            \frac{\Pr(\model{0})}{\Pr(\model{1})}.
\end{align}
The Bayes Factor represents the posterior odds of model $\model{0}$ and model $\model{1}$ given a measured \ce{^14CO} profile. $\Pr(\model{0})$ and$\Pr(\model{1})$ are the prior probabilities of the models. Assuming no reason to favor either model {\sl a priori}, the Bayes Factor reduces to the likelihood ratio
\begin{align}
  B_{01} &= \frac{\Pr(\ce{^14CO}|\model{0})}{\Pr(\ce{^14CO}|\model{1})}
    = \frac{\int d\vec{\theta}_0\ \Pr(\ce{^14CO}|\vec{\theta}_0,\model{0}) \Pr(\vec{\theta}_0|\model{0})}{\int d\vec{\theta}_1\ \Pr(\ce{^14CO}|\vec{\theta}_1,\model{1}) \Pr(\vec{\theta}_1|\model{1})},
\end{align}
where $\vec{\theta}_0$ and $\vec{\theta}_1$ describe the parameters of models 0 and 1. For example, if model 0 corresponds to the assumption of constant cosmic-ray flux, $\vec{\theta}_0$ would be the muon production rates $\vec{\theta}_0=\qty[P_{0,\mu^-},\ P_{0,\mu_f}]$; and if model 1 corresponds to a cosmic ray flux that varies linearly in time, then $\vec{\theta}_1=\qty[P_{0,\mu^-}(0),\ P_{0,\mu_f}(0),\ a]$ where
\begin{align}
  P_{0,\mu^-}(t) &= P_{0,\mu^-}(0)\qty[1 + a\cdot t] &
  &\text{and} &
  P_{0,\mu_f}(t) &= P_{0,\mu_f}(0)\qty[1 + a\cdot t].
\end{align}
In our calculation, we use uninformative uniform priors for all model parameters; for example
\begin{align}
  \Pr(P_{0,\mu_f}|\model{0}) &= \frac{1}{P_{0,\mu_f}^\text{max}-P_{0,\mu_f}^\text{min}} = \frac{1}{\Delta P_{0,\mu_f}},
  &
  \Pr(a|\model{1}) &= \frac{1}{a_\text{max}-a_\text{min}} = \frac{1}{\Delta a}.
\end{align}
This makes the calculation of the sensitivity conservative, though it can be updated to incorporate experimental constraints on the parameters. The likelihoods $\Pr(\ce{^14CO}|\vec{\theta}_i,\model{i})$ assume uncorrelated 3\% Gaussian uncertainties $\sigma_j$ on the \ce{^14CO} data, a conservative estimate of the measurement uncertainties. Thus, the likelihood of the constant flux model $\model{0}$ is
\begin{multline}
  \Pr(\ce{^14CO}|\model{0}) = \int dP_{0,\mu^-}\int dP_{0,\mu_f}\
  \frac{1}{\Delta P_{0,\mu^-}} \cdot \frac{1}{\Delta P_{0,\mu_f}}\\
  \prod_{j=1}^N \frac{1}{\sqrt{2\pi}\sigma_j}
  \exp{-\frac{1}{2}\qty(\frac{\ce{^14CO}_j - c(z_j|P_{0,\mu^-}, P_{0,\mu_f})}{\sigma_j})^2},
\end{multline}
where $c(z_j|P_{0,\mu^-}, P_{0,\mu_f})$ is the expected \ce{^14CO} concentration at ice equivalent depth $z_j$, given $P_{0,\mu^-}$ and $P_{0,\mu_f}$. Similar expressions can be derived for the likelihoods $\Pr(\ce{^14CO}|\model{1})$ of models where the cosmic-ray flux (and thus, the \ce{^14CO} production rate) varies as a function of time.

\begin{figure}[ht]
  \centering
  \includegraphics[width=\textwidth]{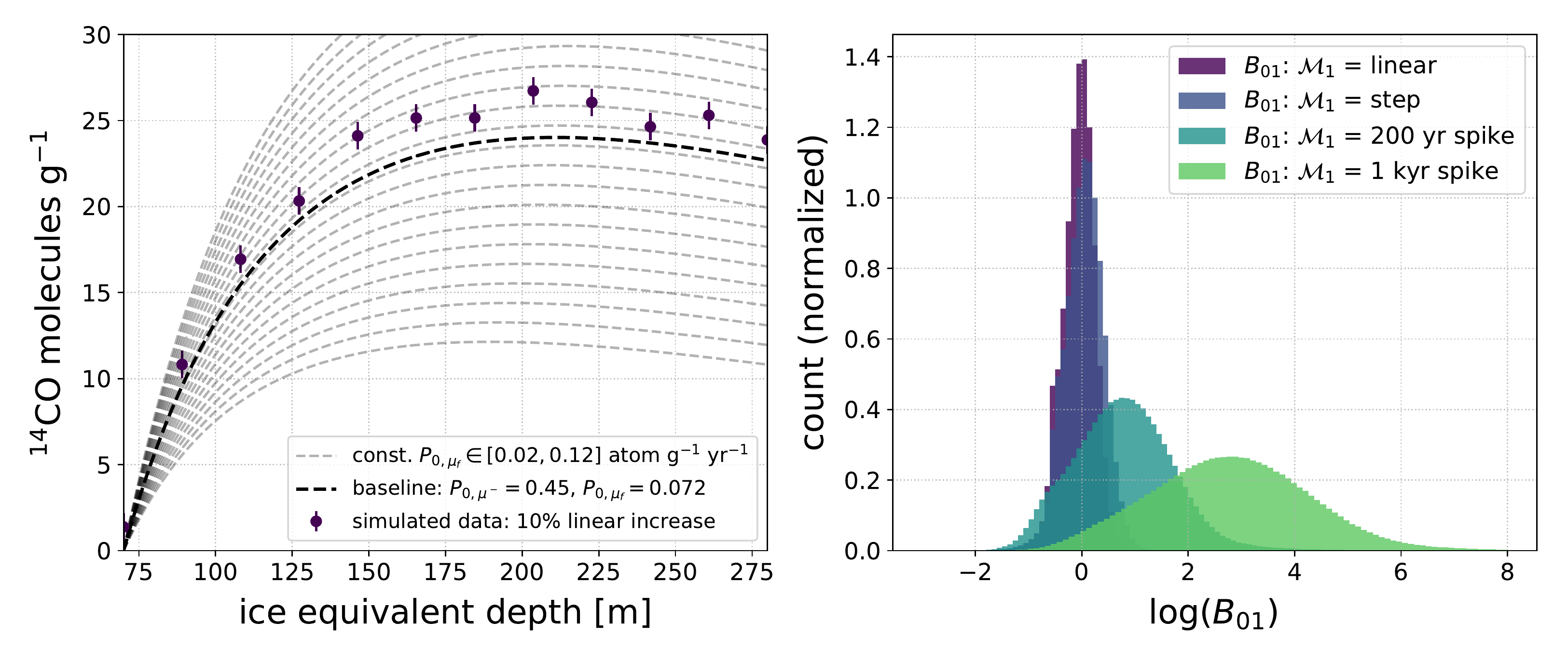}
  \caption{{\sl Left:} the constant-rate ``baseline'' profile based on measurements of $P_{0,\mu^-}$ and $P_{0,\mu_f}$ at Taylor Glacier (thick dashed line) and other selected constant-rate profiles (thin dashed lines). A simulated \ce{^14CO} measurement in which $P_{0,\mu_f}$ increases linearly by 10\% is indicated by the data points.
  {\sl Right:} distribution of the Bayes Factor $B_{01}$ (see eq.~\ref{eq:bf}) for simulated data sets from the baseline profile. Four time-varying models --- a linear increase in the production rates, an abrupt step-like increase, and impulsive increases --- are chosen for the alternative model $\model{1}$.  By the construction of $B_{01}$, values of $\log{B_{01}}>0$ favor the constant-rate model $\model{0}$.}
  \label{fig:co14_data_bf}
\end{figure}

The Bayes Factor describes the posterior odds of favoring model 0 over 1. The interpretation of the odds ratio conventionally follows the scale suggested by Jeffreys \cite{Jeffreys:1998} and Kass and Raftery \cite{Kass:1995}, with $B_{01}>10^2$ considered decisive evidence in favor of model 0, and $B_{01}<10^{-2}$ considered decisive evidence in favor of model 1. While the Bayesian interpretation can be quite useful, in this work we convert $B_{01}$ to a frequentist test statistic to estimate the sensitivity of the \ce{^14CO} profile at Dome C to a time-varying flux of cosmic rays. We calibrate $B_{01}$ as follows:
\begin{enumerate}
  \setlength{\itemsep}{-0.25em}
  \item We produce $3.5\times10^6$ random realizations of the \ce{^14CO} profile assuming constant production rates of $P_{0,\mu^-}=0.45$ \molgyr and $P_{0,\mu_f}=0.072$ \molgyr, and conservative measurement uncertainties of $\sigma_j/\ce{^14CO}_j\approx3\%$ in the \ce{^14CO} vertical profile.
  \item For each time-varying flux under consideration, we compute the distribution of $B_{01}$ using the random constant-flux data sets. We denote these distributions $B_{01}^{\model{0}}$. The distributions corresponding to four time-varying models are shown in the right panel of Fig.~\ref{fig:co14_data_bf}.
  \item We next produce $10^4$ random realizations of models $\model{1}$ with time-varying cosmic ray fluxes, and compute the Bayes Factor $B_{01}^*$ for each. We expect that $B_{01}^*$ will be, on average, much smaller than $B_{01}^{\model{0}}$ since $\model{1}$ will be favored. The left panel of Fig.~\ref{fig:co14_data_bf} shows such a simulated data set.
  \item For each computed $B_{01}^*$, we compute a frequentist $p$-value using the distribution $B_{01}^{\model{0}}$. I.e.,
  \[
    p = \Pr(B_{01}^{\model{0}} < B_{01}^*|\model{0})
  \]
  gives the tail probability that a constant-flux model produces a Bayes Factor smaller than a time-varying flux model. In fact, this is effectively the chance probability that a constant flux can be mistaken for a time-varying flux due to a statistical fluctuation.
  \item Finally, we report the sensitivity as the value of the parameters $\vec{\theta}_1$ for which at least 50\% of the $10^4$ simulated data sets yield $p\lesssim10^{-3}$ --- a ``$3\sigma$'' result --- and at least 50\% of simulated data sets yield $p\lesssim2.9\times10^{-7}$ --- a ``$5\sigma$'' result.
\end{enumerate}
Note that the choice of the baseline \ce{^14CO} profile for $\model{0}$ causes the distribution of $B_{01}$ to peak near 1 for linear and step-like models (see Fig.~\ref{fig:co14_data_bf}). Thus many realizations of the constant-rate model will result in $B_{01}\ll1$, producing a conservative estimate of the sensitivity.

\begin{table}[ht]
  \centering
  \begin{tabular}{|l|c|c|}
    \hline
    \multirow{2}{*}{\bf Difference from Baseline Model} & \multicolumn{2}{c|}{\bf Sensitivity} \\
    & \multicolumn{1}{c}{$\quad\mathbf{3\sigma}$ {\bf ($\geq$50\% of trials)}$\quad$} & \multicolumn{1}{c|}{$\quad\mathbf{5\sigma}$ {\bf ($\geq$50\% of trials)}$\quad$}
    \\
    \hline
    \rowcolor{gray!20}
    Linear Increase over 7 kyr
    & 14\% & 21\%
    \\
    Step-like Increase at 3.5 kyr
    & 9\% & 15\%
    \\
    \rowcolor{gray!20}
    Impulsive Increase (200 yr)
    & 90\% & 152\%
    \\
    Impulsive Increase (1 kyr)
    & 17\% & 30\%
    \\
    \hline
  \end{tabular}
  \caption{Simulated ssensitivity to changes in the depth-concentration profile of \ce{^14CO} at Dome C, computed using a Bayes Factor $B_{01}$ calibrated to the null hypothesis of a constant \ce{^14CO} production rate. Sensitivities are reported as the magnitude of time-varying changes to the cosmic-ray flux which yields a statistical significance of at least $3\sigma$ ($p\lesssim10^{-3}$) and at least $5\sigma$ ($p\lesssim3\times10^{-7}$) in at least 50\% of simulated data sets.}
  \label{tab:sensitivity}
\end{table}

The sensitivity to several time-varying models is listed in Table~\ref{tab:sensitivity}. For a model $\model{1}$ in which the slow and fast muon production rates increase linearly with time, the calibrated Bayes Factor produces a $3\sigma$ difference from the constant-flux model $\model{0}$ at least $50\%$ of the time when the rate of increase is 14\%. At the $5\sigma$ level, the analysis is sensitive to linear increases of 21\% in the production rates (and by proxy, the cosmic ray flux). These are large changes, but are well within the current 30\% uncertainty in the past flux of cosmic rays.

We note that the Bayes Factor is more sensitive to step-like changes in the \ce{^14CO} production rates in the middle of our 7~kyr measurement interval, largely because the shape of the predicted concentration profile differs more dramatically from models which assume a constant flux (see the right panel of Fig.~\ref{fig:co14_profile_models}). We estimate that measurements will be sensitive to an abrupt increase in the flux of 9\% at the $3\sigma$ level, and 15\% at the $5\sigma$ level.

We also explore the sensitivity to impulsive changes in the flux of cosmic rays and the \ce{^14CO} production rates, simulating ``spikes'' in the rate which last 200 and 1,000 years, respectively. In both cases, the impulse is simulated in the center of the 7~kyr accumulation interval of \ce{^14CO} at Dome C. We find that the analysis is sensitive only to relatively large (and likely unphysical) impulsive changes in $P_{0,\mu^-}$ and $P_{0,\mu_f}$. As one might expect, the sensitivity increases in proportion to the duration of the impulse.  

\section{Conclusion}\label{sec:conclusion}

Measurements of the historical flux of cosmic rays, and searches for time variability in the flux, are relevant for many biological, climatological, and geological studies. We expect that the flux of Galactic cosmic rays at Earth is not constant over time due to the effects of supernovae and the motion of the solar system through regions of low and high density in the interstellar medium. Detailed records of radionuclides indicate that the flux of cosmic rays below 100~GeV has varied significantly during the past $10^4$~yr. Since cosmic rays in this energy range are modulated by the solar cycle, these variations can be used to study historical changes in properties of the Sun and heliosphere if we assume a constant flux in the background of Galactic cosmic rays.

Currently available data suggest that the Galactic cosmic ray flux has remained approximately constant on timescales as long as $10^6$ years, but with an uncertainty of 30\% or more. Measurements of \ce{^14CO} in polar ice cores from low snow accumulation sites promise a newly accurate method of searching for time variability in the historical cosmic-ray flux above the energy range where solar modulation plays a large effect. Using recent measurements of \ce{^14CO} production by slow and fast muons in ice, we estimate that at a site like Dome C we can constrain abrupt changes in the flux of high-energy cosmic rays to 15\% at the $5\sigma$ level, and linear changes to 21\% at $5\sigma$. We point out that in our calculations, we made conservative assumptions about the measurement uncertainties of \ce{^14CO}, our prior knowledge of the production rates $P_{0,\mu^-}$ and $P_{0,\mu_f}$, and the choice of ``null'' hypothesis of constant flux. Thus, reduced measurement uncertainties or improved knowledge of production rates in the future will increase the sensitivity of this analysis.

\section{Acknowledgments}

This work has been made possible by support from the University of Rochester, the U.S. National Science Foundation Office of Polar Programs, the David and Lucile Packard Foundation, and the Australian Nuclear Science and Technology Organization.

\bibliographystyle{ICRC}
\bibliography{references}

\end{document}